# Transmission zeros and ultrasensitive detection in complex systems

Yuhao Kang and Azriel Z. Genack

Queens College and The Graduate Center of the City University of New York

Flushing, NY 11367, USA

Understanding vanishing transmission in Fano resonances in quantum systems and metamaterials, and ultralow transmission in disordered media is key to elucidating optical interactions. Using analytic theory and numerical simulations, we discover the topological structure and motion under deformation of transmission zeros in the complex energy plane. We demonstrate the zeros can be controlled to create ultranarrow Lorentzian lines in spectra of transmission time and diverging rates of frequency change of transmission zeros as they converge. This suggests a new approach to ultrasensitive detection.

There has been longstanding interest in understanding the suppression of scattering in quantum and classical systems. The steep asymmetric drop to zero in spectra of Fano resonances arises from the interference of a narrow mode and a continuum or broad mode [1–3], while ultralow transmission in the lowest transmission eigenchannel of the transmission matrix (TM) of multichannel disordered media [4–12] is due to interference of far-off-resonance modes [8].

Fano's analysis [1] was introduced to explain inelastic electron scattering in helium but has been applied beyond atomic physics to nuclear and condensed matter physics, electronics, and optics [2,13]. The TM was developed to explain the scaling of conductance in the quasi-1D wire geometry [4] but has been intensively studied recently because it allows the control of transmission of classical waves [8,9,14,10,15,12]. By properly forming the incident wave, it is possible to achieve perfect transmission, with transmission eigenvalue $\tau_1 = 1$ of the matrix $tt^\dagger$, where $t$ is the $N \times N$ TM [4,8,15]. The average of the lowest transmission eigenchannel $\tau_N$ is $e^{-2L/\ell}$, where $L$ is the sample length and $\ell$ is the transport mean free path [4–7]. Whether transmission can be identically zero has remained an open question. Since the average of $\tau_N$ increases with absorption, it is possible that even weak absorption would thwart the possibility of a null in transmission [10,16].

The increasing power of nanofabrication and the continuing discovery of novel properties of waves in metamaterial has heightened interest in exploiting singularities in the SM [17–19] or in portions of the SM, such as the reflection matrix [20], for applications to sensing, switching, lasing and energy deposition [17–21]. The singularities of the SM in unitary systems are complex conjugate pairs of poles and zeros in the complex energy or frequency plane. Incident radiation is completely absorbed when a zero of the SM is brought to the real axis. Such coherent perfect absorption (CPA) [17,22–27] is the time reversal of an outgoing wave at the lasing threshold. Lasing and CPA may occur simultaneously in a PT-symmetric system in which a pole and its conjugate zero are brought to the real axis together [28,29].

Zero reflection is also achieved in any subset of input channels when zeros of the reflection matrix (RM) are on the real axis [20,30]. The reflection zeros may be found anywhere in the complex plane. The reflection time difference (RTD) between the two sides of a quasi-1D sample, can be expressed as a sum of Lorentzians associated with the reflection zeros [31,32]. This is a counterpoint to the Wigner time delay, which can be expressed as the sum of resonances corresponding to Lorentzians functions for the poles [33].

The poles of the TM are the same as for the SM and RM, but the zeros which correspond to interference of the transmitted field are entirely different. It is therefore of interest to explore the structure of the zeros of the TM in the complex plane. This would give deeper insight into Fano resonances and low transmission eigenchannels and might allow their control for diverse applications. Possible correlation between the transmission zeros could facilitate their approach and serve as a pathway to high sensitivity of the transmission zeros to perturbation of a medium. That is analogues to the heightened sensitivity that arises when two poles converge as they approach an exceptional point (EP) [34,35].

In this Letter, we demonstrate the topological structure of zeros of the TM. In a unitary medium, the transmission zeros appear either as single zeros on the real axis or as complex conjugate pairs. The single zeros are topologically constrained to move on the real axis with deformation of the sample, while two single zeros and a conjugate pair of zeros may interconvert when they meet. The rate of frequency change of zeros near this zero point (ZP) diverges so that changes in the sample can in principle be detected with ultrahigh resolution. In a unitary system, the transmission time is the density of states (DOS), which is a sum of Lorentzian lines associated with the poles or resonances. But, in the presence of loss or gain, the zeros are manifest in the spectrum of transmission time as Lorentzian lines with linewidths which vanishes as the zeros are brought to the real axis. The prospects for ultrasensitive detection of perturbations near a ZP are discussed.

The TM $t$ is a quadrant of the SM $S = \begin{bmatrix} r & t' \\ t & r' \end{bmatrix} = \begin{bmatrix} V & \\ & U \end{bmatrix} \begin{bmatrix} -\sqrt{1-\tau} & \sqrt{\tau} \\ \sqrt{\tau} & \sqrt{1-\tau} \end{bmatrix} \begin{bmatrix} V^\dagger & \\ & U^\dagger \end{bmatrix}$. Here, $\tau$ is the diagonal matrix of transmission eigenvalues, and $U$ and $V$ are unitary matrices [12]. In 1D, the transmission time is the spectral derivative of the phase of the transmitted field $\tau_T = \frac{d \arg(t)}{dE}$, where $\hbar = 1$ and $E$ is the angular frequency, $\omega$, for classical waves [36–40]. Generalizing this relation to quasi-1D gives $\tau_T = \frac{d}{dE} \arg\det(t)$. We show in Supplemental Material section I [41], that, in a lossless medium, this relation gives $\tau_T = \pi\rho$, where $\rho$ is the density of states (DOS) [42,43].

To separate out the impact of resonances and zeros upon the transmission time, we derive the relation between $\tau_T$ and the DOS using the Heidelberg model [33,44–47], in which the SM is expressed as $S = \frac{I - iK}{I + iK}$, where $K = \pi W^\dagger \frac{1}{E - H_{in}} W$. Here, $H_{in}$ is the internal Hamiltonian of the scattering region and *W* is the coupling matrix between the channels in the leads and the modes within the medium. The coupling between the *N* channels in the leads and the *M* quasi-normal modes of the system within the spectral range of interest is given via the matrix $W = [W_1 \ W_2]$, where the $M \times N$ matrix $W_{1/2}$ couples the scattering region and its surroundings. The expression for the determinant of the TM is obtained in Supplemental Material section II [41],

$$\det(t) = (-2\pi i)^N \frac{\det(E - H_{in})\det(W_2^\dagger \frac{1}{E - H_{in}} W_1)}{\det(E - H_{eff})}, \quad (1)$$

where $H_{eff} = H_{in} - i\pi W W^\dagger$ is the effective Hamiltonian of the scattering region.

The numerator of Eq. (1) can be expressed as $\prod_{i=1}^{M-N}(E - \eta_i)$, where the $\eta_i = Z_i + i\zeta_i$ denote the zeros of $\det(t)$ (Supplemental Material section II) [41], while the denominator is given by $\det(E - H_{eff}) = \prod_n (E - \lambda_n)$, where $\lambda_n = E_n - i\gamma_n$ are the poles of the resonances, and $\gamma_n$ is the halfwidth of the mode. This yields the factorized expression,

$$\det(t) \sim \frac{\prod_{i=1}^{M-N}(E - \eta_i)}{\prod_{j=1}^{M}(E - \lambda_j)}. \quad (2)$$

Since the transmission time is given by $\tau_T = \frac{d}{dE}\arg\det(t)$, the sum of the contributions from poles and zeros is

$$\tau_T = \tau_p + \tau_z = \sum_n \frac{\gamma_n}{(E - E_n)^2 + \gamma_n^2} + \sum_i \frac{\zeta_i}{(E - Z_i)^2 + \zeta_i^2}. \quad (3)$$

Since $\prod_{i=1}^{M-N}(E - \eta_i)$ is always real for real *E* in a unitary system (Supplemental Material section II [41]), the $\eta_i$ must be disposed symmetrically with respect to the real axis. The zeros are therefore either real or conjugate pairs. Therefore $\tau_z$ vanishes in a unitary system and $\tau_T$ is due solely to the poles, which is the DOS [33,43].

The symmetry of the $\eta_i$ is broken by loss or gain so that $\tau_z$ no longer vanishes. For a system with uniform internal loss or gain, $H_{in} = H_0 - i\gamma$, the position of the zeros of $\det(t)$ shifts down by $i\gamma$, giving, $\eta_i' = \eta_i - i\gamma$. The contribution of a single zero to $\tau_T$ is $\tau_z = \frac{-\gamma}{(E - Z)^2 + \gamma^2}$, with peak value $-\gamma^{-1}$. For a conjugate pair of zeros in a unitary medium, $\eta = Z \pm i\zeta$, the corresponding zeros for the nonunitary system are at $\eta' = Z + i(\pm\zeta - \gamma)$. A pair of zeros then contribute to the transmission time with, $\tau_z = \frac{-\zeta-\gamma}{(E-Z)^2+(-\zeta-\gamma)^2} + \frac{\zeta-\gamma}{(E-Z)^2+(\zeta-\gamma)^2}$, giving a local extremum at $E=Z$ of $\frac{2\gamma}{\zeta^2-\gamma^2}$. In general, $\tau_z$ does not vanish in a nonunitary system, however, it can vanish in a 1D PT-symmetric system [28,48] with balanced loss and gain (Supplemental Material section IV [41]).

We first explore the impact of transmission zeros on the transmission and transmission time in a random quasi-1D sample. We carry out simulations for a quasi-1D system using the tight-binding model simulated with the open-source package Kwant [49] for a sample with $N = 5$. The onsite energy is 4 in the uniform leads and distributed randomly within the scattering region over [4-w, 4+w], with w=1.1, and nearest neighbor coupling $-1$ are shown in Fig. 1. The profiles of energy density of transmission eigenchannels with perfect and vanishing transmission in a lossless sample are shown in Fig. 1a for points in the spectra of the transmission eigenvalues indicated in Fig 1b.

The inset in Fig. 1c shows that $\tau_T$ and $\tau_p$ coincide in the lossless system, with $\tau_z = 0$. With uniform internal loss of $\gamma = 0.002$, however, $\tau_T$ and $\tau_p$ differ, as shown in the spectrum of $\tau_z = \tau_T - \tau_p$ in Fig. 1c. The imaginary coordinates of the transmission zeros can be determined from the extrema in $\tau_z$. The peak of $\tau_z$ indicated by the red circle in Fig. 1c shows that one of the zeros of a pair is slightly above the real axis. The value at the peak, $\frac{2\gamma}{\zeta^2-\gamma^2}$ gives $\zeta = 5.3\times10^{-3}$. Thus, the upper transmission zero of the pair at $E = 0.666$ would be moved to the real axis by adding loss of $\gamma = 5.3\times10^{-3}$. At this energy, the lowest transmission eigenchannel vanishes, $\tau_N = 0$, as indicated by the dip in $\tau_N$ in Fig. 1d.

Simulation of transmission in the lossless system does not allow a definitive determination of whether the lowest transmission eigenchannel is identically zero. This can be determined, however, from the depth of the dip of $\tau_z$ when absorption is added. For a single real transmission zero in the lossless system, $\tau_z$ would dip to $\gamma^{-1} = 500$, as is indeed found in Fig. 1c. If $\det(t)$ at this frequency in the lossless sample were merely exponentially small, there would be a pair of conjugated zeros close to the real axis, which would give a dip of $2\gamma^{-1} = 1000$. The dip of 500 allows us to conclude that the smallest transmission eigenvalue at the blue circle is identically zero in the lossless case.

There are three transmission zeros within the energy range of Fig. 1. Their positions are shown in Fig. 1e as the upper zero of the pair is moved to the real axis. Both transmission zeros and poles are topological phase singularities. The phase of $\det(t)$ can be seen in the phase map of Fig. 1f to increase (decrease) by $2\pi$ in a counterclockwise rotation about the transmission zeros (poles), which are indicated by red (blue) squares. The phase singularities correspond to conserved topological charges of +1 for each zero, and are analogous to the phase singularities in the speckle pattern of scattered waves [50–52].

We now consider the motion of transmission zeros with displacement of an element of the sample. The sample is a lossy billiard with reflecting disks coupled to its surroundings via two single-channel leads. The lowest disk is displaced, as shown in Fig. 2a. The position of the poles are found using COMSOL mode solver. Figure 2b shows the zeros of a pair approaching each other on a curved trajectory as the lowest disk is displaced to the right. The zeros remain equidistant from the line $\zeta = -i\gamma$ along the trajectory and meet on the line at a ZP. At the ZP, the phase changes by $4\pi$ in a counterclockwise circuit of the zeros so that the topological charge is conserved as the paired zeros are transformed to single zeros, as can be seen in Fig. 2c. With further displacement of the disk, two single zeros appear and move along the line $\zeta = -i\gamma$. It is noteworthy that the poles seen in Fig. 2c hardly moves in the deformation in which a pair of zeros is converted to two single zeros.

One of the zeros of a pair can be brought to the real axis in an absorbing medium by deforming the sample, as indicated by the star in Fig. 2b. The corresponding intensity profile and energy flow is shown in Fig. 2a. As was the case for the quasi-1D sample, transmission in the billiard can also be made to vanish by adding absorption (Supplemental Material section V [41]). The vanishing of transmission in an absorbing sample is easily perturbed since even a small perturbation moves the zero off the real axis. In contrast, the single zero in a Fano resonance in a lossless system cannot be eliminated unless two zeros meet and are converted to a pair of zeros. This is essentially the time reversal of the scenario in Fig. 2b.

The first frame in Fig. 3a shows the evolution of two zeros in the complex plane as the lowest disk moves horizontally. The diverging slopes of the trajectories of the single and paired zeros relative to the $x$-coordinate of the disk at the ZP indicates the diverging sensitivity of the zeros to at the ZP. Near the ZP at $Z_0 = 18.2148$ GHz , $\zeta_0 = 0$ for $x_0 = 2.003$ cm, $|Z - Z_0| = \alpha_Z \sqrt{x - x_0}$, with $\alpha_Z = 0.17$ for the single zeros moving along the real axis, and $|\zeta - \zeta_0| = \alpha_\zeta \sqrt{x - x_0}$ with $\alpha_\zeta = 0.18$ for the conjugate pair moving perpendicular to the real axis. The gives the square root singularity in the sensitivities of $Z$ and $\zeta$ around the ZP: $dZ/dx = \alpha_Z/2\sqrt{x - x_0}$ and $d\zeta/dx = \alpha_\zeta/2\sqrt{x - x_0}$. A comparison between transmission spectra in which the zeros are at (red) or near (blue) the ZP is shown in the inset. A displacement of the lowest disk of 0.001 cm produces a 0.01 GHz shift between the transmission dips, which is a fractional shift of $5 \times 10^{-4}$. Figure 3b shows the sensitivity of the zeros relative to change of the radius of the lowest disk for which the sensitivity of $Z$ and $\zeta$ have a square root divergence at the ZP with $\alpha_Z = 0.11$ and $\alpha_\zeta = 0.09$. Thus, an easily resolved separation between the transmission dips is produced by a fractional change of $\sim 10^{-3}$ of the diameter of the disk. This translates to a layer of thickness 0.2 nm for a disk of 200 nm in a scaled structure. Since the dips are clearly resolved, a thickness change which is a small fraction of an atomic diameter could be detected.

In summary, we have shown that the transmission and transmission time in complex systems are determined by the zeros as well as the poles of the wave equation. In addition to sharp dips in transmission, ultranarrow Lorentzian lines of transmission time can be created by positioning a zero slightly off the real axis. The zeros and poles are topological phase singularities and determine the map of the phase of $\det(t)$ in the complex energy plane for a given structure. Unlike for the SM or RM, there are strong topological constraints on the positions of zeros of the TM in the complex plane. This requires that the zeros either lie on the real axis of the complex energy plane or are conjugate pairs. These principles govern the structure and evolution of zeros, but many fundamental questions remain. These include the ratio of the number of single and paired zeros and the distribution of the imaginary coordinate of the paired zeros in diverse structured and random systems. The distribution of zeros of the SM has been calculated in chaotic cavities [53]. The ability to control the position of zeros by adding loss or gain or by deforming the sample, combined with the diverging sensitivity of zeros near a ZP provides a powerful platform for monitoring structural changes in a medium. Extreme sensitivity is also found near EPs [18,35,54]. But the high sensitivity of transmission zeros near a ZP does not require the precise tuning of a sample so that the modes of the medium coalesce [18,35,54–56], and can be achieved even in Hermitian systems with appropriate modification of the sample. Optical measurements of the frequency difference between zeros in transmission or between peaks or dips in transmission time or phase derivative near a ZP in would allow monitoring of perturbations on a subatomic scale.

# Figures

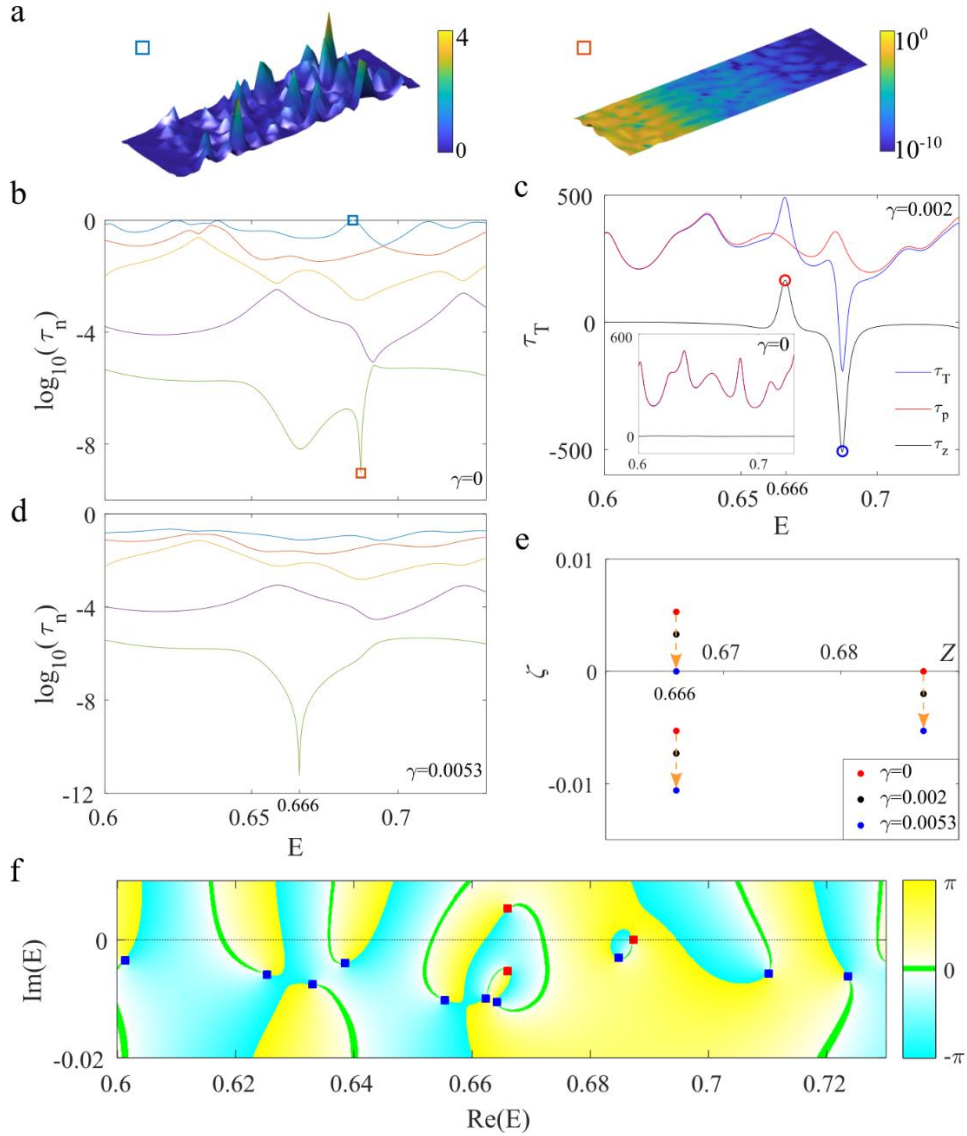

**Fig. 1. Transmission zeros in a random quasi-1D sample.** Simulations are carried out in a sample of width 20 and length 60 which supports 5 channels in the energy range of the simulation. (a) Intensity profile for the perfectly transmitting eigenchannel at $E = 0.6845$ and the eigenchannel with vanishing transmission at $E = 0.6873$, corresponding, respectively, to the blue and red squares, in Fig. 1b. The colorbar is linear for the first pattern and logarithmic for the second pattern. (b) Spectra of the five transmission eigenvalues in the lossless system. (c) Transmission times for on-site loss $\gamma = 0.002$. $\tau_p$ is calculated by integrating the imaginary part of the local Green's function $\tau_p = -\int \mathrm{Im}\, G(r,r,E)d\vec{r}$. $\tau_z = \tau_T - \tau_p$ is given by the black curve. The inset shows that $\tau_p$ and $\tau_T$ coincide, with $\tau_z = 0$ in the lossless system. (d) Lossy system with $\gamma = 0.0053$ in which one transmission zero lies on the real axis at $E = 0.666$ with $\tau_N = 0$. (e) Displacement of three transmission zeros in the complex plane for the losses in (b-d). (f)

Phase map of $\arg \det(t)$ in the complex energy plane. Red (blue) squares indicate transmission zeros (poles).

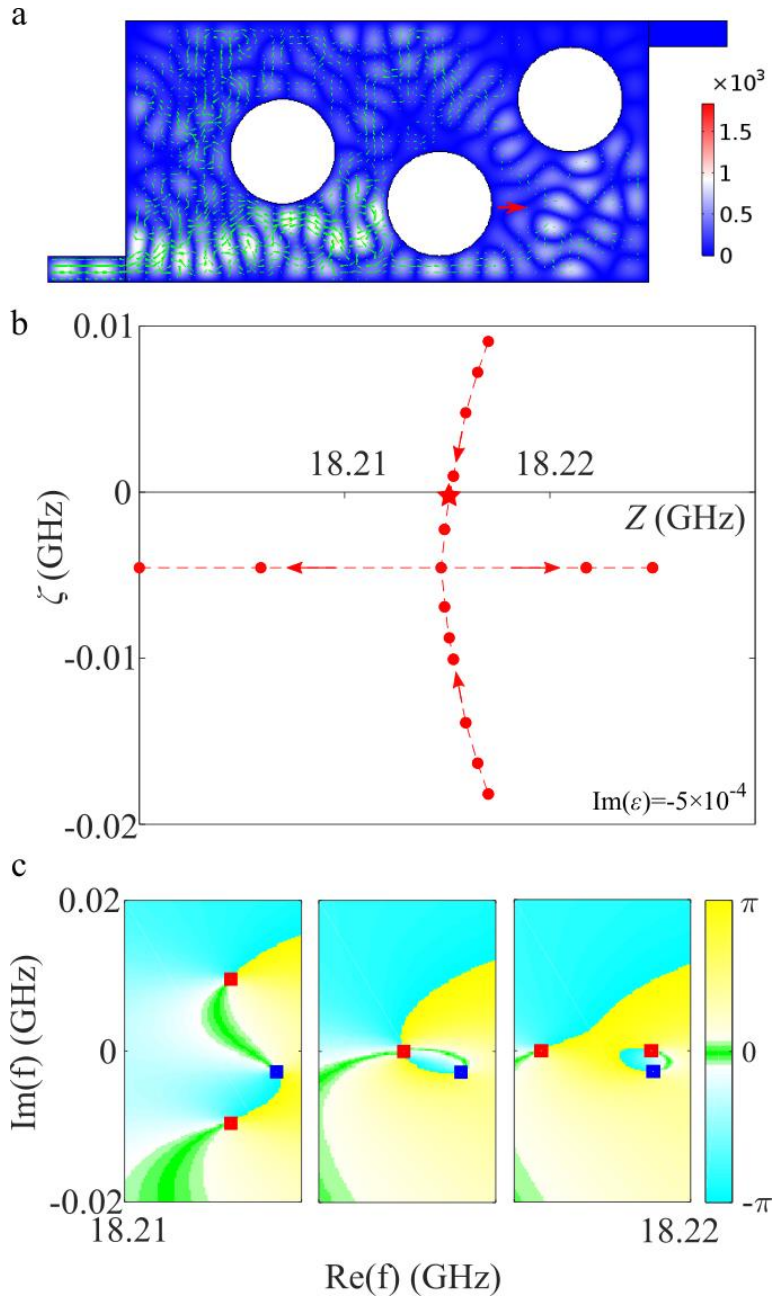

**Fig. 2. Interconversion of single and paired transmission zeros in a lossy billiard.** (a) Profiles of energy flow and field amplitude at the transmission zero indicated by the star in (b). The length of the arrows indicates the logarithm of the flux. The amplitude of the field is given by the colorbar on the right. The permittivity in the system is $\varepsilon = 1 - 5 \times 10^{-4} i$. The sample width and length are 10 and 20 cm. The leads have a width of 1 cm. The diameter of the disk is 4 cm. (b) Trajectories of two transmission zeros with x-coordinates of the center of the lowest disk at

1.996, 1.998, 2, 2.002, 2.0024, 2.0028, 2.003, 2.005, 2.008 (cm). The two degenerate zeros meet when *x*=2.003 cm. The star indicates the zero-transmission mode in the lossy system. (c) Phase diagram of transmission between 18.21 and 18.22 GHz when the lowest disk is at 2.000 cm (left), at 2.003 cm (middle) and 2.0033 cm (right). Red (blue) squares indicate transmission zeros (poles).

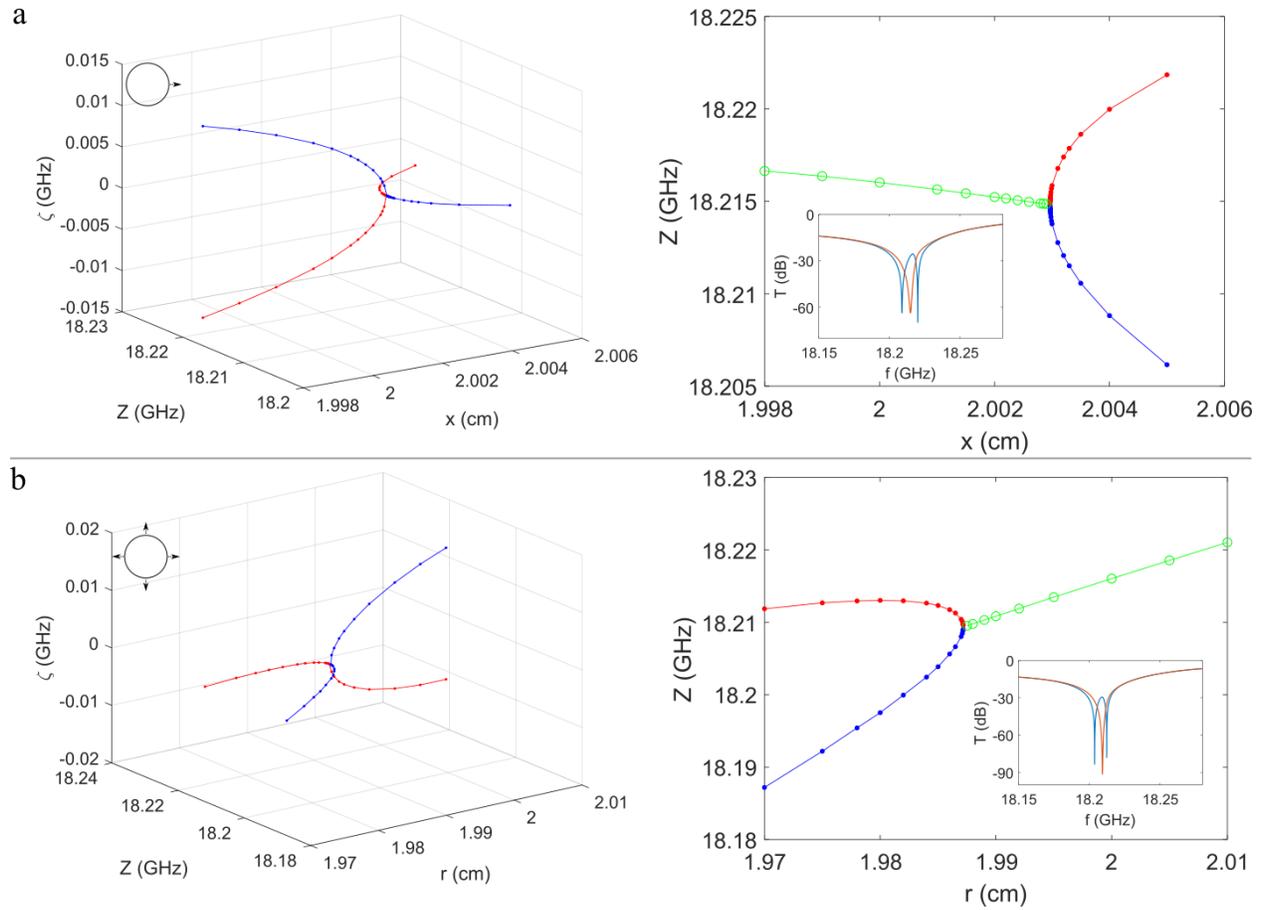

**Fig. 3. Sensitivity of near degenerate transmission zeros in lossless billiard.** (a) Left panel shows the trajectory of two zeros relative to the coordinate of the lowest disk. Right panel shows the variation of the frequency of the zeros with displacement of the disk. The green circles and the curve drawn through them gives the real frequencies of the pair of zeros which meet at *x*=2.003 cm. The trajectories for the two real zeros after they are created at *x*~2.003 cm are shown in red and blue. The red (blue) curve in the inset of the right panel shows the transmission spectrum when the x-coordinate of the lowest disk is 2.003 (2.004) cm. (b) The trajectory of two transmission zeros with radius *r* of the lowest disk with center at *x*=2cm. Two curves meet at *r*=1.9872 cm. The inset shows transmission spectra for a radius of the disk of 1.9872 cm (red) and 1.9850 cm (blue).

**Acknowledgements**

We thank Yiming Huang for sharing results of simulations and for valuable discussions. This work is supported by the National Science Foundation under EAGER Award No. 2022629 and by PSC-CUNY under Award No. 63822-00 51.

**Supplemental Material for**

**Transmission zeros and ultrasensitive detection in complex systems**

Yuhao Kang and Azriel Z. Genack

Queens College and The Graduate Center of the City University of New York

Flushing, NY 11367, USA

## I. Relationship between the transmission time and the DOS in a lossless system

The SM in lossless systems can be expressed as $S = \begin{bmatrix} V & \\ & U \end{bmatrix} \begin{bmatrix} -\sqrt{1-\tau} & \sqrt{\tau} \\ \sqrt{\tau} & \sqrt{1-\tau} \end{bmatrix} \begin{bmatrix} V^\dagger & \\ & U^\dagger \end{bmatrix}$,

where U and V are unitary matrices. The DOS can then be calculated via the Wigner-Smith matrix

$$\rho = -i\frac{1}{2\pi}TrS^\dagger \frac{dS}{dE} = -i\frac{1}{\pi}Tr(U^\dagger \frac{dU}{dE} - V^\dagger \frac{dV}{dE}). \tag{S1}$$

The transmission time for a multichannel system is

$$\tau_T = \frac{d}{dE}\arg\det(t), \tag{S2}$$

Since $t = U\sqrt{\tau}V^\dagger$ and $\sqrt{\tau}$ is a real diagonal matrix,

$$\arg\det(t) = \arg\det(U) - \arg\det(V). \tag{S3}$$

U can be written as $U = e^{iH}$ since U is unitary, where H is a Hermitian matrix.

$\det(U) = \det(e^{iH}) = e^{iTr(H)}$. Since $Tr(H)$ is a real number, the phase derivative of $\det(U)$ can be written as

$$\begin{aligned}\frac{d}{dE}\arg\det(U) &= \frac{d}{dE}Tr(H) = Tr(\frac{d}{dE}H) \\ &= -iTr(\frac{d}{dE}\log(U)) = -iTr(U^{-1}\frac{dU}{dE}) \\ &= -iTr(U^\dagger \frac{dU}{dE}).\end{aligned} \tag{S4}$$

Similarly,

$$\frac{d}{dE}\arg\det(V) = -iTr(V^{\dagger}\frac{dV}{dE}), \tag{S5}$$

Combining Eqs. (S1-S5) gives the transmission time in a quasi-1D system,

$$\tau_{T} = -iTr(U^{\dagger}\frac{dU}{dE}) + iTr(V^{\dagger}\frac{dV}{dE}) = \pi\rho, \tag{S6}$$

which is the summation of eigenchannel transmission times proposed in Ref. [1].

## II. Determinant of the transmission matrix

The Heidelberg model is built on the connection between the scattering process and the resonant modes of the system. Suppose the system supports $M$ quasi-normal modes and is connected to its surroundings with $N$ incoming and $N$ outgoing channels. The coupling matrix is $W = [W_1 \ W_2]$, where $W_{1/2}$ is an $M \times N$ matrix. Usually $M \gg N$. The SM can be expressed as [2,3]

$$S(E) = I - 2\pi i W(E)^{\dagger} \frac{1}{E - H_{eff}(E)} W(E), \tag{S7}$$

where the effective Hamiltonian is

$$\begin{aligned} H_{eff} &= H_{in} + \int dE' \frac{W(E')W^{\dagger}(E')}{E - E'} \\ &= H_{in} - i\pi W W^{\dagger} + P\int dE' \frac{W(E')W^{\dagger}(E')}{E - E'}. \end{aligned} \tag{S8}$$

Usually, the principal value integral is ignored [2]. We assume the coupling matrix is independent of energy in the following derivation.

Since $(A_{n \times n} + B_{n \times m} C_{m \times n})^{-1} B_{n \times m} = A_{n \times n}^{-1} B_{n \times m}(I_m + C_{m \times n} A_{n \times n}^{-1} B_{n \times m})^{-1}$ [4,5], we find

$$\frac{1}{E - H_{eff}} W_1 = (E - H_{in} + i\pi W_1 W_1^{\dagger} + i\pi W_2 W_2^{\dagger})^{-1} W_1 = A^{-1} W_1 (I + i\pi W_1^{\dagger} A^{-1} W_1)^{-1}, \tag{S9}$$

where $A = E - H_{in} + i\pi W_2 W_2^{\dagger}$.

Eq. (S7) gives the TM

$$t = -2\pi i W_2^\dagger \frac{1}{E - H_{eff}} W_1 \qquad (S10)$$
$$= -2\pi i W_2^\dagger A^{-1} W_1 (I + i\pi W_1^\dagger A^{-1} W_1)^{-1}.$$

The determinant of the TM is then

$$\det(t) = \frac{\det(-2\pi i W_2^\dagger A^{-1} W_1)}{\det(I + i\pi W_1^\dagger A^{-1} W_1)}$$

$$= \frac{\det(-2\pi i W_2^\dagger A^{-1} W_1)}{\det(I + i\pi A^{-1} W_1 W_1^\dagger)}$$

$$= \frac{\det(A)\det(-2\pi i W_2^\dagger A^{-1} W_1)}{\det(A + i\pi W_1 W_1^\dagger)}$$

$$= \frac{\det(E - H_{in} + i\pi W_2 W_2^\dagger)\det(-2\pi i W_2^\dagger \frac{1}{E - H_{in} + i\pi W_2 W_2^\dagger} W_1)}{\det(E - H_{eff})}.$$

The second equality makes use of Sylvester's determinant theorem
$$\det(I_m + A_{m\times n} B_{n\times m}) = \det(I_n + B_{n\times m} A_{m\times n}).$$

The numerator of $\det(t)$ can be simplified as follows:

$$\det(E - H_{in} + i\pi W_2 W_2^\dagger)\det(-2\pi i W_2^\dagger \frac{1}{E - H_{in} + i\pi W_2 W_2^\dagger} W_1)$$

$$= \det(E - H_{in})\det(I + i\pi \frac{1}{E - H_{in}} W_2 W_2^\dagger)\det(-2\pi i W_2^\dagger \frac{1}{E - H_{in} + i\pi W_2 W_2^\dagger} W_1)$$

$$= \det(E - H_{in})\det(I + i\pi W_2^\dagger \frac{1}{E - H_{in}} W_2)\det(-2\pi i W_2^\dagger \frac{1}{E - H_{in} + i\pi W_2 W_2^\dagger} W_1)$$

$$= \det(E - H_{in})(-2\pi i)^N \det(W_2^\dagger \frac{1}{E - H_{in} + i\pi W_2 W_2^\dagger} W_1 + i\pi W_2^\dagger \frac{1}{E - H_{in}} W_2 W_2^\dagger \frac{1}{E - H_{in} + i\pi W_2 W_2^\dagger} W_1)$$

$$= \det(E - H_{in})(-2\pi i)^N \det[W_2^\dagger (I + i\pi \frac{1}{E - H_{in}} W_2 W_2^\dagger) \frac{1}{E - H_{in} + i\pi W_2 W_2^\dagger} W_1]$$

$$= \det(E - H_{in})\det(-2\pi i W_2^\dagger \frac{1}{E - H_{in}} W_1).$$

To summarize, we obtain two expressions for $\det(t)$:

$$\det(t) = (-2\pi i)^N \frac{\det(E - H_{in} + i\pi W_2 W_2^\dagger)\det(W_2^\dagger \frac{1}{E - H_{in} + i\pi W_2 W_2^\dagger} W_1)}{\det(E - H_{eff})}, \qquad (S11)$$

and

$$\det(t) = (-2\pi i)^N \frac{\det(E - H_{in})\det(W_2^\dagger \frac{1}{E - H_{in}} W_1)}{\det(E - H_{eff})}.$$ (S12)

We define the numerators of $\det(t)$ as $Y_1 // Y_2$ in the two expression above,

$$Y_1 = \det(E - H_{in} + i\pi W_2 W_2^\dagger)\det(W_2^\dagger \frac{1}{E - H_{in} + i\pi W_2 W_2^\dagger} W_1),$$

and

$$Y_2 = \det(E - H_{in})\det(W_2^\dagger \frac{1}{E - H_{in}} W_1),$$

$E$ is defined over the whole complex plane.

$Y_2$ is ill-defined at the eigenvalues $t_j$ of the Hamiltonian of the closed system, $H_{in}$. When $H_{in}$ is Hermitian, the $t_j$ are real. $Y_1$ is the analytic continuation of $Y_2$, which is well-defined at $E = t_j$.

$Y_2$ can be expressed as

$$\begin{aligned} Y_2 &= \det(E - H_{in})\det[\frac{W_2^\dagger adj(E - H_{in})W_1}{\det(E - H_{in})}] \\ &= \frac{\det[W_2^\dagger adj(E - H_{in})W_1]}{\det(E - H_{in})^{N-1}} \\ &\propto \frac{\prod_i (E - s_i)}{[\prod_j (E - t_j)]^{N-1}}. \end{aligned}$$ (S13)

For every factor involving $t_j$ in the denominator of Eq. (S13), there must be $N-1$ factors involving $s_i$ in the numerator; otherwise, $Y_2$ would diverge as $E$ approaches $t_j$. However, $Y_1$ indicates that $E$ is well-defined along the entire real axis.

Since the highest order of $\det[W_2^\dagger adj(E - H_{in})W_1]$ is $(M-1)N$ and the highest order of $\det(E - H_{in})$ is $M$, the number of zeros of Eq. (S13) is $(M-1)N - M(N-1) = M - N$.

We conclude that the numerator of $\det(t)$ can be factorized as $Y = \prod_i (E - \eta_i)$, $i = 1 \sim (M - N)$ and $\eta$ is the set difference between $s$ and $t$, $\{\eta\} = \{s\} \setminus \{t\}$. Equation (S12) can therefore be rewritten as $\det(t) \sim \dfrac{\prod_{i=1}^{M-N}(E - \eta_i)}{\prod_{j=1}^{M}(E - \lambda_j)}$, where the $\lambda_j$ an eigenvalue of $H_{eff}$.

Similarly, for transmission from the right to left sides,

$$\det(t') = (-2\pi i)^N \det(E - H_{eff})^{-1} \det(E - H_{in}) \det(W_1^\dagger \dfrac{1}{E - H_{in}} W_2).$$

In a reciprocal system, $t^T = t'$, so that $\det(W_2^\dagger \dfrac{1}{E - H_{in}} W_1) = \det(W_1^\dagger \dfrac{1}{E - H_{in}} W_2)$. When, in addition, the system does not possess internal loss or gain, $H_{in}^\dagger = H_{in}$. When $E$ is on the real axis, $(W_2^\dagger \dfrac{1}{E - H_{in}} W_1)^\dagger = W_1^\dagger \dfrac{1}{E - H_{in}} W_2$. Combining these equations, we find for a unitary reciprocal system that $\det(W_2^\dagger \dfrac{1}{E - H_{in}} W_1)$ is real for real $E$. Thus, the numerator of Eq. (S12) is real for real $E$.

### III. Interference between a single mode and a continuum

The Heidelberg model only considers the contribution of resonances. But in systems in which waveguides are coupled along their length to localized modes, the transmission includes a direct path not associated with the resonances of the medium. We consider a spectral region in which transmission is due to a single isolated mode and a slowly varying background. Suppose the complex frequency of the mode is $E_0 - i\dfrac{\Gamma}{2}$, where $\Gamma = \Gamma_1 + \Gamma_2 + \Gamma_a$, $\Gamma_1$ ($\Gamma_2$) is the linewidth due to the coupling of the mode to left (right) lead and $\Gamma_a$ is due to loss or gain and is positive (negative) when the system possesses loss (gain). Based on coupled-mode theory, the transmitted field can be expressed as $t = b + \dfrac{\sqrt{\Gamma_1 \Gamma_2}}{i(E_0 - E) + \dfrac{\Gamma}{2}} e^{i(\varphi + \varphi_b)}$, where

$b$ represents the background field, and the second term on the right is the contribution of the resonance. $\varphi$ is the difference between the phases of the resonance and background, $\varphi_b$ is the phase of background $b$. We focus on the rapid phase change due to the resonance and set $b$ to be real with $\varphi_b = 0$. In general, $\dfrac{d\varphi_b}{dE}$ contributes to the time delay.

The expression for $\varphi$ is given in Ref. [6], $\cos\varphi = -\dfrac{(\Gamma_1 + \Gamma_2)b}{2\sqrt{\Gamma_1\Gamma_2}}$. This gives,

$\arg(t) = \arctan(\dfrac{\Delta E + E_0 - E}{\Gamma_a / 2}) - \arctan(\dfrac{E_0 - E}{\Gamma / 2})$, where $\Delta E = \pm\sqrt{\dfrac{\Gamma_1\Gamma_2}{b^2} - \dfrac{(\Gamma_1 + \Gamma_2)^2}{4}}$. The transmission time,

$$\tau_T = \dfrac{\Gamma/2}{(E-E_0)^2 + (\Gamma/2)^2} - \dfrac{\Gamma_a/2}{(E-E_0-\Delta E)^2 + (\Gamma_a/2)^2}, \qquad (S14)$$

is the sum of a term due to the mode $\tau_p$ and an additional term. The additional term is proportional to the strength of the loss or gain. We will show that the additional term is equal to $\tau_z$ in Eq. (3) of the main text.

The single zero in the system has a complex value

$\eta = Z + i\zeta = E_0 - i\left(\dfrac{\sqrt{\Gamma_1\Gamma_2}}{b}e^{i\varphi} + \dfrac{\Gamma}{2}\right) = E_0 + \Delta E - i\dfrac{\Gamma_a}{2}$. So that

$\tau_z = \dfrac{\zeta}{(E-Z)^2 + \zeta^2} = \dfrac{-\Gamma_a/2}{(E-E_0-\Delta E)^2 + (\Gamma_a/2)^2}$.

When $\Gamma_a = 0$, $\tau_T$ equals $\pi\rho$. The resultant Fano resonance exhibits an abrupt phase jump of $-\pi$. But, aside from this singularity, the phase derivative is still a sum of Lorentzian functions. The relative contribution of the term associated with the zero in Eq. (S14) can be estimated from the ratio $\dfrac{\tau_z}{\tau_p}$ at the energy $E = E_0$, $\dfrac{\tau_z}{\tau_p} = \dfrac{\Gamma_a \Gamma}{\Gamma_a^2 + \dfrac{4\Gamma_1\Gamma_2}{b^2} - (\Gamma_1 + \Gamma_2)^2}$. To better understand the denominator, we rewrite $t$ as $t = t\frac{-iq-i\delta}{1-i\delta}$, where $\delta = \frac{E-E_0}{\Gamma/2}$.

Comparing this expression to $t = b + \frac{\sqrt{\Gamma_1\Gamma_2}}{i(E_0-E)+\frac{\Gamma}{2}}e^{i\varphi}$, gives $iq = -\frac{2\sqrt{\Gamma_1\Gamma_2}}{\Gamma b}e^{i\varphi} - 1$. Thus,

$|q|^2 \Gamma^2 = \Gamma_a^2 + \frac{4\Gamma_1\Gamma_2}{b^2} - (\Gamma_1+\Gamma_2)^2$ and $\left|\frac{\tau_z}{\tau_p}\right| = \frac{1}{|q|^2}\frac{|\Gamma_a|}{\Gamma}$. $q$ is the asymmetry factor of the Fano resonance in a unitary system. When $|q|^2 < \frac{|\Gamma_a|}{\Gamma}$, the point of zero transmission, $E_0 - \frac{\Gamma}{2}q$, is close to the central frequency of the mode $E_0$. In this case, $\tau_T < 0$. This model of a simple system shows that the transmission time can be negative in systems with internal loss when the spectrum has a point of zero transmission near the mode. However, when $|q|^2 > \frac{|\Gamma_a|}{\Gamma}$, the spectrum is nearly Lorentzian and $\tau_z$ is negligible relative to the contribution of the resonant term.

**IV. Spectrum of the transmission time**

In a lossless or a 1D PT-symmetric system, poles and zeros of the SM form conjugate pairs, $\det(S) = \frac{\prod_n(E-\lambda_n^*)}{\prod_n(E-\lambda_n)}$, where $\lambda_n = E_n - i\gamma_n$. Thus, the Wigner time is given by

$$\tau_w = \frac{\partial}{\partial E}\arg(\det(S)) = -\sum_n \frac{2\mathrm{Im}(\lambda_n)}{(E-\mathrm{Re}(\lambda_n))^2 + \mathrm{Im}(\lambda_n)^2} = \sum_n \frac{2\gamma_n}{(E-E_n)^2 + \gamma_n^2}. \quad (S15)$$

Supplementary Note I shows that $\tau_w = 2\tau_T$ in a lossless system. This relation still holds in a 1D PT-symmetric system as shown in the following. The SM can be written as

$S = \begin{bmatrix} r & t' \\ t & r' \end{bmatrix} = t\begin{bmatrix} \frac{1-|t|^{-2}}{ic} & 1 \\ 1 & ic \end{bmatrix}$, where $c = -ir'/t$ [7]. The phase difference between reflection from two sides is then 0 or $\pi$ depends on whether $|t|$ larger or smaller than 1. So that the Wigner delay time is $\tau_w = -i\frac{\partial}{\partial E}\ln\det(S) = 2\frac{\partial\varphi_t}{\partial E} = 2\tau_T$.

Thus, in a lossless or a 1D PT-symmetric system, $\tau_T = \sum_n \frac{\gamma_n}{(E-E_n)^2 + \gamma_n^2}$.

### V. Transmission and transmission time near a zero

We now examine the way the displacement of transmission zeros in the complex energy plane by absorption affects the transmission and transmission time in the billiard shown in Fig. 2. The modes of the system are found with use of the harmonic inversion method [8]. The fit of a sum of modes to spectra of transmission and phase of the transmitted field in the lossy billiard is shown in Fig. S1a. The sum of Lorentzian lines associated with poles gives the spectrum of $\tau_p$ shown in Fig. S1b. As for the random quasi-1D sample, the narrow peak in $\tau_z$ at $f = 18.216$ GHz shows that one of a pair of zeros is slightly above the real axis.

The transmission zero can be brought to the real axis by adding loss. The position of the upper zero of the pair of $\zeta = 0.0098$ GHz in the lossless system is found from the peak in $\tau_z$. The decay rate that corresponds to $\gamma = \zeta$ that is needed to bring the zero to the real axis in a system with unit real part of the index of refraction is $\gamma = \zeta = f n_i = \frac{f \text{Im}(\varepsilon)}{2}$, or

$\text{Im}(\varepsilon) \approx -2\zeta / f = -1.1 \times 10^{-3}$.

Transmission spectra as the position of a zero crosses the real axis by adding loss is shown in the transmission spectra shown in Fig. S1c. Transmission decreases with increasing loss until it vanishes when the zero reaches the real axis. After that, transmission increases as the pair of zeros move down away from the real axis. The energy flow at the point of zero transmission for $\text{Im}(\varepsilon) = -1.1 \times 10^{-3}$ is shown in Fig. S1d. The intensity in the outgoing lead vanishes.

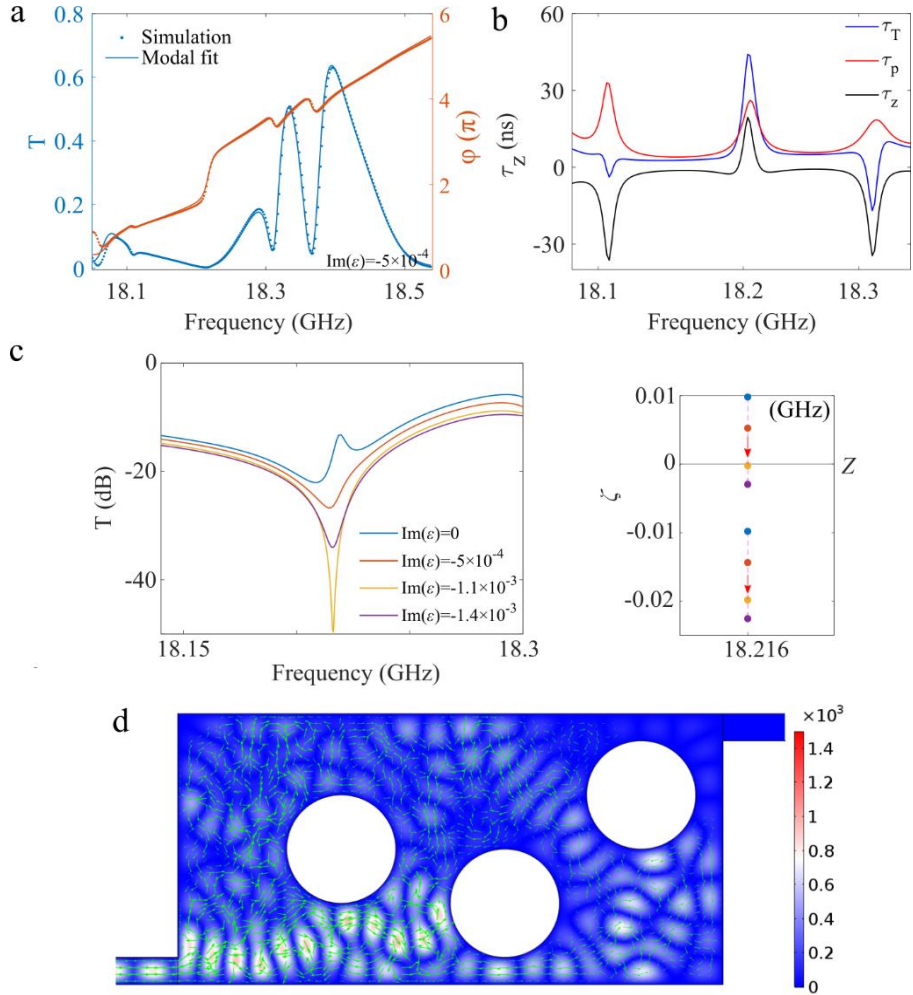

**Fig. S1.** We consider the structure in Fig. 2a with the x-coordinate of the center of the lowest disk at 2 cm. (a) The permittivity in the system is the same as in Fig. 2a. The blue and red curves are the transmission and phase, respectively. COMSOL simulations are shown as the dots, while the fit using modes found using the harmonic inversion method are shown as continuous curves. (b) Spectra of $\tau_T$, $\tau_p$ and $\tau_z$. $\tau_p$ is the sum of Lorentzian lines of the modes extracted with use of harmonic inversion in (a). The spectrum of $\tau_z$ is obtained from the difference between $\tau_T$ and $\tau_p$. The peak of $\tau_z$ at 18.216 GHz and 19.5 ns shows that one transmission zero is still above the real axis, while the two dips indicate that two real zeros in the lossless system are pushed below the real axis. (c) Transmission spectra for different values of loss. The right panel shows $\zeta$ for different values of $\text{Im}(\varepsilon)$. One zero of the pair of zeros at frequency 18.216 GHz hits the real axis for $\text{Im}(\varepsilon) = -1.1 \times 10^{-3}$. (d) Profiles of energy flow and field amplitude at the transmission zero.